\AddToHook{package/hyperref/after}{\RequirePackage{orcidlink}}
\def\TetraStateAuthorBlock{%
  \IEEEauthorblockN{Junyu Huo\,\orcidlink{0009-0003-9278-2870}}%
  \IEEEauthorblockA{\textit{Phanvic}\\
    Calgary, Alberta, Canada\\
    junyu@phanvic.ca}%
  \and
  \IEEEauthorblockN{Zihao Wan\,\orcidlink{0009-0004-4389-2763}}%
  \IEEEauthorblockA{\textit{Phanvic}\\
    Calgary, Alberta, Canada\\
    vic@phanvic.ca}%
  \and
  \IEEEauthorblockN{Ziqi Mao\,\orcidlink{0009-0009-4422-1079}}%
  \IEEEauthorblockA{\textit{University of Calgary}\\
    Calgary, Alberta, Canada\\
    sarah.mao@ucalgary.ca}%
  \and
  \IEEEauthorblockN{Gouri Ginde\,\orcidlink{0000-0001-7519-3503}}%
  \IEEEauthorblockA{\textit{University of Calgary}\\
    Calgary, Alberta, Canada\\
    gouri.ginde@ucalgary.ca}%
}
\def\TetraStateThanks{We thank Harper Shi (Queen's University) for feedback
  on revisions of this paper and help with its references. }

\documentclass[10pt,conference]{IEEEtran}

\usepackage{amsmath}
\usepackage{amssymb}
\usepackage{booktabs}
\usepackage{graphicx}
\usepackage{microtype}
\usepackage{xcolor}
\usepackage{url}
\usepackage[hidelinks]{hyperref}\definecolor{findingframe}{RGB}{112,48,160}\definecolor{findingfill}{RGB}{246,240,251}\newcommand{\finding}[1]{\par\medskip\noindent\fcolorbox{findingframe}{findingfill}{\parbox{\dimexpr\columnwidth-2\fboxsep-2\fboxrule\relax}{\small #1}}\par\medskip}
\usepackage{tikz}
\usetikzlibrary{arrows.meta,positioning,fit}

\newcommand{\CosmosPairs}{506}
\newcommand{\CosmosTasks}{2,024}
\newcommand{\CosmosClassified}{405}
\newcommand{\CosmosCoverage}{80.0\%}
\newcommand{\CosmosAccuracy}{70.6\%}
\newcommand{\CosmosHallucinations}{43}
\newcommand{\CosmosHallRate}{10.6\%}

\newcommand{\CosmosTies}{90}
\newcommand{\CosmosAbstainPairs}{11}
\newcommand{\CosmosPfourCount}{25}

\newcommand{\CosmosStable}{175}
\newcommand{\CosmosFragile}{111}
\newcommand{\CosmosUnstable}{76}
\newcommand{\CosmosHallLow}{7.9\%}
\newcommand{\CosmosHallHigh}{13.5\%}

\newcommand{\VjepaTasks}{2,024}
\newcommand{\VjepaClassified}{470}
\newcommand{\VjepaCoverage}{92.9\%}
\newcommand{\VjepaAccuracy}{48.5\%}
\newcommand{\VjepaHallucinations}{204}
\newcommand{\VjepaHallRate}{43.4\%}

\newcommand{\VjepaTies}{36}

\newcommand{\VjepaStable}{190}
\newcommand{\VjepaFragile}{38}
\newcommand{\VjepaUnstable}{38}
\newcommand{\VjepaHallLow}{38.9\%}
\newcommand{\VjepaHallHigh}{47.9\%}

\newcommand{\VjepaAblationVerdictChanges}{415}
\newcommand{\VjepaAblationPairChanges}{214}
\newcommand{\VjepaExactClassified}{442}
\newcommand{\VjepaExactCoverage}{87.4\%}
\newcommand{\VjepaExactAccuracy}{51.1\%}
\newcommand{\VjepaExactHallRate}{35.3\%}

\newcommand{\VjepaArchivedClassified}{470}
\newcommand{\VjepaArchivedCoverage}{92.9\%}
\newcommand{\VjepaArchivedAccuracy}{48.3\%}
\newcommand{\VjepaArchivedHallRate}{44.0\%}
\newcommand{\PaddingAccuracyDelta}{0.2}
\newcommand{\PaddingPairChanges}{11}
\newcommand{\PaddingReversals}{8}
\newcommand{\VjepaAblationAccuracyDelta}{2.6}
\newcommand{\VjepaAblationStableToFragile}{35}
\newcommand{\VjepaAblationNewHallucinations}{16}
\newcommand{\VjepaAblationCoverageDrop}{5.5}

\newcommand{\PaddingRegressed}{5}
\newcommand{\PaddingImproved}{6}
\newcommand{\PaddingStableToHall}{3}
\newcommand{\VjepaAblationRegressed}{114}
\newcommand{\VjepaAblationImproved}{100}
\newcommand{\VjepaAblationStableLost}{74}
\newcommand{\VjepaAblationSameModal}{69}
\newcommand{\VjepaModalCorrect}{228}
\newcommand{\VjepaExactModalCorrect}{226}
\newcommand{\VjepaArchivedModalCorrect}{227}
\newcommand{\VjepaAblationHiddenRegressions}{43}
\newcommand{\VjepaAblationUnstableToHall}{8}

\newcommand{\TotalTies}{126}

\newcommand{\ConfigurationSensitive}{263}

\newcommand{\ArtifactURL}{\url{https://anonymous.4open.science/r/tetrastate-artifact-5C1E/}}

\title{One Verdict Is Not Enough: Regression Testing Learned World Models and Their Test Harnesses}

\ifdefined\TetraStateAuthorBlock
  \author{\TetraStateAuthorBlock}
\else
  \author{}
\fi

\begin{document}
\maketitle

\begin{abstract}
Learned world models ship as software components whose model versions, preprocessing, and test harnesses change between releases. Work on model updates already counts per-sample regressions, or negative flips, that an accuracy check hides. Two problems remain: a test can pass under one configuration and fail under another, and a harness fault can produce erroneous scores and wrong verdicts without a crash. We present \emph{TetraState}, a regression-testing protocol for paired possible/impossible video tests. Each test runs under four fixed model-native configurations; its number of correct verdicts assigns one of four outcome classes. Inconclusive tests and failed executions are reported separately alongside the verdict rate. Revisions are compared by per-test outcome transitions, and the harness is versioned as part of the system under test. On all \CosmosPairs{} IntPhys~2 Main tests and two heterogeneous systems, Cosmos~3 Nano and V-JEPA~2, two V-JEPA~2 harness revisions move modal accuracy by only \PaddingAccuracyDelta{} and \VjepaAblationAccuracyDelta{} points yet change \PaddingPairChanges{} and \VjepaAblationPairChanges{} test labels; in the second, accuracy rises although fewer tests are judged correctly, and \VjepaAblationHiddenRegressions{} regressions keep the modal verdict, so a modal-verdict flip check misses them. These results outline a research agenda on configuration-level regression oracles for learned components.
\end{abstract}

\begin{IEEEkeywords}
  regression testing, test oracle, machine-learning testing, world models, configuration-dependent behavior
\end{IEEEkeywords}

\section{Introduction}\label{sec:intro}

Learned world models predict how scenes evolve and now ship as components for video understanding, simulation, autonomous driving, and robotic planning \cite{ha2018worldmodels,hu2023gaia1,nvidia2026cosmos3,assran2025vjepa2}. Like any software component, they evolve: model versions are updated, preprocessing changes, and test harnesses are patched. Software testing treats code, data, model versions, preprocessing, and configuration as coupled parts of an ML system \cite{zhang2022mltesting,riccio2020testing,sculley2015debt}, and practitioners report regression faults when testing evolving DL systems \cite{you2025evolving}. An accuracy-only regression oracle reduces such a revision to one aggregate: benchmark accuracy before versus after the change.

Accuracy is a lossy regression oracle: a revision can regress some tests and repair others while the total barely moves. Model-update work measures negative flips, inputs that the new model gets wrong although the old one got them right, and backward compatibility \cite{bansal2019updates,srivastava2020backward,yan2021pct,xie2021regression}. These measures compare one verdict per input. For learned components, two further gaps remain. First, verdicts are configuration-dependent: the same test can pass with one frame count or context length and fail with another, so one execution cannot establish consistency across settings. Second, the test harness belongs inside the system under test (SUT). A fault in frame sampling, score conversion, or loss aggregation can produce an erroneous but well-formed score, leading to a wrong verdict (a failure) without a crash \cite{humbatova2020faults}. Neither problem is visible in one aggregate or one verdict per test.
\emph{Motivating example.} A manual code audit during our study revealed a fault in our harness for V-JEPA~2, Meta's video model \cite{assran2025vjepa2}: when converting its per-window losses into a score, the harness averaged in zero-valued padding entries. Fixing the fault moved majority-vote accuracy from \VjepaArchivedAccuracy{} to \VjepaAccuracy{}, so an accuracy-based quality gate would have passed both the faulty and the fixed version. Yet \PaddingPairChanges{} test labels changed, \PaddingReversals{} between correct under every configuration and wrong under every configuration. The aggregate therefore conceals changes that warrant per-test inspection.

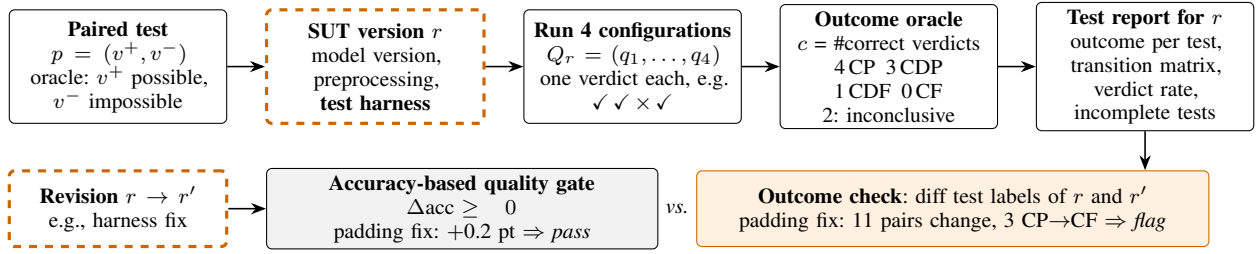
\begin{figure*}[t]
  \centering
  \begin{tikzpicture}[
      font=\footnotesize,
      node distance=5mm,
      box/.style={draw, rounded corners=2pt, align=center, inner sep=2.5pt, minimum height=15mm, text width=27mm},
      flow/.style={-{Stealth[length=2mm]}, semithick},
    ]
    \node[box] (pair) {\textbf{Paired test}\\$p=(v^+,v^-)$\\oracle: $v^+$ possible,\\$v^-$ impossible};
    \node[box, right=of pair, draw=orange!80!black, dashed, very thick] (route) {\textbf{SUT version $r$}\\model version,\\preprocessing,\\\textbf{test harness}};
    \node[box, right=of route] (run) {\textbf{Run 4 configurations}\\$Q_r=(q_1,\ldots,q_4)$\\one verdict each, e.g.\\\checkmark\,\checkmark\,$\times$\,\checkmark};
    \node[box, right=of run] (oracle) {\textbf{Outcome oracle}\\$c$ = \#correct verdicts\\4\,CP\enspace 3\,CDP\\1\,CDF\enspace 0\,CF\\2: inconclusive};
    \node[box, right=of oracle] (report) {\textbf{Test report for $r$}\\outcome per test,\\transition matrix,\\verdict rate,\\incomplete tests};
    \draw[flow] (pair) -- (route);
    \draw[flow] (route) -- (run);
    \draw[flow] (run) -- (oracle);
    \draw[flow] (oracle) -- (report);
    \node[box, below=6mm of pair, minimum height=10mm, draw=orange!80!black, dashed, very thick] (revision) {\textbf{Revision $r \to r'$}\\e.g., harness fix};
    \node[box, right=of revision, text width=50mm, minimum height=10mm, fill=gray!10] (acc) {\textbf{Accuracy-based quality gate}\\$\Delta$acc $\geq 0$\\padding fix: $+$\PaddingAccuracyDelta{} pt $\Rightarrow$ \emph{pass}};
    \node[box, right=of acc, text width=66mm, minimum height=10mm, fill=orange!12, draw=orange!80!black] (diff) {\textbf{Outcome check}: diff test labels of $r$ and $r'$\\padding fix: \PaddingPairChanges{} pairs change, \PaddingStableToHall{} CP$\to$CF $\Rightarrow$ \emph{flag}};
    \draw[flow] (revision) -- (acc);
    \path (acc) -- node[font=\footnotesize\itshape] {vs.} (diff);
    \draw[flow] (report.south) -- (report.south |- diff.north);
  \end{tikzpicture}
  \caption{TetraState as a regression test. Each pair runs under four SUT-native settings (12/24/36/42 span the official V-JEPA~2 context range); an outcome oracle summarizes the verdicts, with harness semantics in the versioned SUT. CP: consistent pass; CDP: configuration-dependent pass; CDF: configuration-dependent fail; CF: consistent fail. Comparing outcomes reveals changes an accuracy delta conceals.}
  \label{fig:overview}
\end{figure*}

We propose \emph{TetraState}, a regression-testing protocol with three elements: each benchmark pair is a test case executed under a fixed set of configurations; an outcome oracle replaces the binary pass/fail verdict; and revisions are compared by per-test outcome transitions, analogous to test-result diffs for code changes (Fig.~\ref{fig:overview}). Here, \emph{state} means an outcome class summarizing a test's correctness profile across the declared configurations. The harness is versioned as part of the SUT. Our contributions are (1) a configuration-level outcome oracle extending per-test flip counts, with explicit verdict rate and a SUT version identifier that includes harness semantics; (2) an open implementation with validation checks for heterogeneous scoring APIs; and (3) emerging evidence on a full benchmark split showing that outcome transitions expose harness revisions that accuracy conceals.

\section{TetraState}\label{sec:solution}

\subsection{Paired Tests and SUT Versions}

For a matched pair $p=(v^+,v^-)$, the benchmark declares $v^+$ physically possible and $v^-$ impossible, e.g., a solid object is blocked by a barrier versus passing through it. This supplies an expected ordering without an exact expected score \cite{barr2015oracle}. A \emph{route} $r$ is a SUT comprising a model, model version, preprocessing, scoring code, and an ordered test configuration set $Q_r=(q_{r,1},\ldots,q_{r,K})$; changing any element creates a distinct SUT version. Under each configuration, an \emph{adapter} scores both videos and selects the more physically possible one; verdict $y_{p,r,k}$ is correct iff it selects $v^+$. Scores are compared only within a pair, SUT, and configuration, so the protocol treats the model as a black box without a shared score scale. Related configurations follow metamorphic and behavioral testing \cite{segura2016survey,ribeiro2020checklist,tian2018deeptest,stevens2025mars}, with verdict invariance as a candidate relation. The paired oracle is known, so configurations probe consistency. They are deterministic, applied symmetrically, and recorded in the SUT identifier; semantic preservation remains an assumption to examine.

\subsection{Outcome Oracle and Verdict Rate}

For four valid verdicts, let $c_{p,r}$ count correct verdicts. The four outcome classes are \emph{consistent pass} (CP), $c=4$; \emph{configuration-dependent pass} (CDP), $c=3$; \emph{configuration-dependent fail} (CDF), $c=1$; and \emph{consistent fail} (CF), $c=0$. A $2$--$2$ split is inconclusive; a failed or abstaining execution yields an error verdict and leaves the test incomplete. Neither receives a decisive outcome class, and their counts are reported separately. The \emph{verdict rate} is the fraction of tests with a decisive outcome. Exact counts avoid a tunable threshold, although the configuration set remains a design choice (Sec.~\ref{sec:instantiation}). Classes describe only the declared settings. Configuration dependence need not indicate a fault: a short frame sample may miss the violating event. CDP and CDF differ from flaky tests, whose outcomes vary across repeated runs of one setting \cite{luo2014flaky,dutta2020flaky}: here executions are deterministic and outcomes vary across configurations, much like configuration-dependent failures in configurable software systems \cite{medeiros2016sampling}. The classes separate this behavior from consistent failures and inconclusive tests.

\subsection{Outcome-Transition Regression Check}

Given reports for a revision $r \to r'$ on the same tests and configuration set, the check computes outcome transitions, with inconclusive $2$--$2$ tests as $c=2$; incomplete tests are reported separately. A test \emph{regresses} when it passes under fewer configurations relative to the fixed paired oracle. Like a newly failing test in code review, this is a signal to inspect, not proof of a fault. The runner retains every configuration-level verdict; outcome classes summarize these per test for triage. The summary loses detail: \checkmark\checkmark\checkmark$\times$ changing to \checkmark\checkmark$\times$\checkmark{} stays CDP although one configuration regressed and another improved. The check also flags newly consistent fails, loss of consistent passes, and a verdict-rate drop. Each accompanies the accuracy delta; inconclusive or incomplete tests never become passes.

\emph{Relation to prior work.} Model-update studies measure negative flips and train models to reduce them \cite{yan2021pct,xie2021regression}; backward-compatibility work tracks the same effect for deployed models \cite{bansal2019updates,srivastava2020backward}. Regression fuzzing generates inputs on which two DL model versions disagree \cite{you2023drfuzz}. These approaches compare one verdict per input. Evaluation code also shifts aggregate scores, as with BLEU and FID implementations \cite{post2018bleu,parmar2022aliased}; here a harness revision changes per-test correctness patterns with weights fixed. IntPhys~2 reports prompt and frame-count sensitivity \cite{bordes2025intphys2}, and MLTE documents model and system evidence \cite{maffey2023mlte}. TetraState makes configuration-dependent behavior a versioned regression check across heterogeneous scoring APIs. Regression oracles thus range from accuracy deltas through per-test majority flips and outcome transitions to full configuration-level verdict diffs. The full diff retains every observed change; TetraState groups changes per test and reports their direction, while retaining the full diff for inspection.

\subsection{SUT Identifiers and Validation Checks}

Each test execution has a provenance record: model and SUT version identifiers, dataset, split, raw scores, configuration evidence, and video identifiers. A build manifest records the resolved harness configuration, source files, and model and artifact hashes \cite{pineau2021reproducibility}. The runner is append-only and resumable; aggregation rejects mixed SUT identifiers and inconsistent test-category labels. Every harness variant receives a distinct SUT version identifier, so a revision cannot silently overwrite its baseline, just as a code change produces a new build.

\label{sec:instantiation}We instantiate $K=4$ configurations labeled 12, 24, 36, and 42, fixed with the state rules and gates before any full-split run. They span the context lengths covered by the official V-JEPA~2 harness run (12 to 42); Cosmos uses the same frame counts. An even $K$ admits 2--2 inconclusive outcomes. These values are one instantiation, not a tuned choice. The two scoring APIs differ: Cosmos~3 Nano's video-reasoning SUT \cite{nvidia2026cosmos3,nvidia2026cosmoscode} samples exactly that many frames uniformly over the full video, endpoints included. Its score $s(v)=\ell(\mathrm{Impossible}\mid v)-\ell(\mathrm{Possible}\mid v)$ uses mean token log probability $\ell$ of each complete label; the lower-scoring video is selected, and equal scores abstain. V-JEPA~2 \cite{assran2025vjepa2,meta2025vjepa2code} instead uses a context length inside the official harness's 48-frame sliding windows \cite{meta2025intphys2code} and selects the lower mean latent-prediction loss. Its official \texttt{max\_context\_mode} combines early-window predictions with exact-$C$ sliding-window losses; disabling it creates a distinct exact-context SUT version. Validation checks verify Cosmos frame counts and endpoints and V-JEPA~2 model and artifact identities. Labels order settings within a SUT and do not imply equal information across SUTs.

\section{Emerging Results}

\noindent\begin{minipage}{\columnwidth}We ask two questions.\par\noindent\textbf{RQ1:} \emph{Do outcome-transition checks flag harness revisions that an accuracy-based quality gate would accept?}\par\noindent\textbf{RQ2:} \emph{What potential triage signals do outcome classes provide beyond accuracy?}\end{minipage}\par

\emph{Setup.} The entire labeled IntPhys~2 Main split \cite{bordes2025intphys2} has 1,012 videos from 253 scenes, forming \CosmosPairs{} tests across four test categories (input partitions P1--P4, corresponding to physical principles). The loader groups videos by official scene and generation identifiers, requires one possible and one impossible candidate, and rejects missing or unmapped entries. Ground truth comes only from official metadata; inclusion is independent of model output. Each SUT ran \CosmosTasks{} test executions (\CosmosPairs{} tests $\times$ 4 configurations). Because each scene yields two tests, intervals use 10,000 percentile bootstrap replicates resampling scenes.

\begin{table}[t]
  \caption{V-JEPA~2 harness revisions with weights, videos, and tests fixed. Dec.: decisive tests; Corr.: correct modal verdicts; VR: verdict rate; Acc.: modal accuracy; CF: consistent-fail rate (Acc. and CF over decisive tests); Chg.: labels differing from official max. Exact-context accuracy rises while correct tests fall.}
  \label{tab:ablation}
  \centering
  \footnotesize
  \begin{tabular}{@{}lrrrrrr@{}}
    \toprule
    Harness & Dec. & Corr. & VR & Acc. & CF (\%) & Chg. \\
    \midrule
    Archived (padded) & \VjepaArchivedClassified{} & \VjepaArchivedModalCorrect{} & \VjepaArchivedCoverage{} & \VjepaArchivedAccuracy{} & \VjepaArchivedHallRate{} & \PaddingPairChanges{} \\
    Official max & \VjepaClassified{} & \VjepaModalCorrect{} & \VjepaCoverage{} & \VjepaAccuracy{} & \VjepaHallRate{} & -- \\
    Exact context & \VjepaExactClassified{} & \VjepaExactModalCorrect{} & \VjepaExactCoverage{} & \VjepaExactAccuracy{} & \VjepaExactHallRate{} & \VjepaAblationPairChanges{} \\
    \bottomrule
  \end{tabular}
\end{table}

\begin{figure}[t]
  \centering
  \includegraphics[width=\columnwidth]{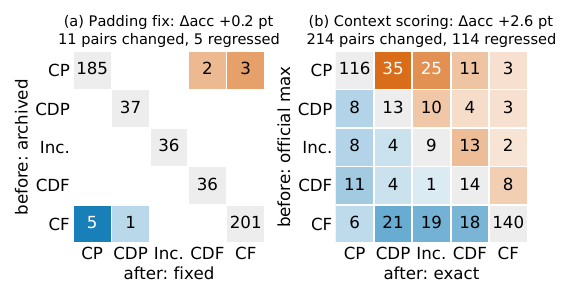}
  \caption{Outcome transitions of two V-JEPA~2 harness revisions. Rows give test labels before and columns after, ordered by correct configurations (4 to 0); Inc. denotes a 2--2 inconclusive test. Cells right of the diagonal (orange) regressed, those left (blue) improved, and diagonal cells (gray) kept their class; shading is relative within each panel.}
  \label{fig:transitions}
\end{figure}

\subsection{RQ1: Regression View of Harness Revisions}\label{sec:regression}

Table~\ref{tab:ablation} and Fig.~\ref{fig:transitions} compare three V-JEPA~2 harness versions with weights, videos, and tests fixed. The \emph{padding fix} repairs a real harness fault found by code audit; we treat it retrospectively as a regression test case. It changes 41/\VjepaTasks{} verdicts and 11/\CosmosPairs{} pair labels, while accuracy moves \PaddingAccuracyDelta{} points and verdict rate is unchanged. \PaddingRegressed{} tests regress, including \PaddingStableToHall{} CP$\to$CF, and \PaddingImproved{} improve; the effects nearly cancel in the aggregate.

The \emph{context-scoring revision} is an intentional behavioral change, not a bug fix: it sets \texttt{max\_context\_mode} to false. Whoever approves it should still know which tests change and in which direction. It changes \VjepaAblationVerdictChanges{}/\VjepaTasks{} verdicts and \VjepaAblationPairChanges{}/\CosmosPairs{} test labels. Aggregate metrics suggest improvement: modal accuracy rises by \VjepaAblationAccuracyDelta{} points and the CF rate falls from \VjepaHallRate{} to \VjepaExactHallRate{}. Both rates use decisive tests, whose count falls from \VjepaClassified{} to \VjepaExactClassified{}; tests with a correct modal verdict actually drop from \VjepaModalCorrect{} to \VjepaExactModalCorrect{}. The outcome check flags \VjepaAblationNewHallucinations{} newly consistent fails, \VjepaAblationStableLost{} of \VjepaStable{} CP tests losing consistency (\VjepaAblationStableToFragile{} still modally correct), and a \VjepaAblationCoverageDrop{}-point verdict-rate drop. In total \VjepaAblationRegressed{} tests regress and \VjepaAblationImproved{} improve. A modal-verdict flip check misses \VjepaAblationSameModal{} of the \VjepaAblationPairChanges{} changes, including \VjepaAblationHiddenRegressions{} regressions (\VjepaAblationStableToFragile{} CP$\to$CDP, \VjepaAblationUnstableToHall{} CDF$\to$CF). The full configuration-level diff retains all \VjepaAblationVerdictChanges{} verdict flips; the outcome diff groups them into \VjepaAblationPairChanges{} test-label changes with a direction. An accuracy-based quality gate that only requires accuracy not to decrease accepts both revisions without inspecting any test.\finding{\textbf{Answer to RQ1:} Two harness revisions that a non-decreasing-accuracy quality gate accepts change \PaddingPairChanges{} and \VjepaAblationPairChanges{} test labels. In the second, \VjepaAblationHiddenRegressions{} regressions keep the modal verdict, so a modal-verdict flip check misses them.}

\subsection{RQ2: Potential Triage Signals Beyond Accuracy}

\begin{table}[!ht]
  \caption{Primary SUT outcome counts in Fig.~\ref{fig:transitions} order. Inc.: inconclusive 2--2 tests; Abst.: incomplete tests with abstentions; VR: verdict rate. CF is a count here and a percentage in Table~\ref{tab:ablation}.} \label{tab:routes} \centering \small \setlength{\tabcolsep}{5pt} \begin{tabular}{@{}lrrrrrrr@{}} \toprule SUT & CP & CDP & Inc. & CDF & CF & Abst. & VR \\ \midrule Cosmos~3 Nano & \CosmosStable{} & \CosmosFragile{} & \CosmosTies{} & \CosmosUnstable{} & \CosmosHallucinations{} & \CosmosAbstainPairs{} & \CosmosCoverage{} \\ V-JEPA~2 max & \VjepaStable{} & \VjepaFragile{} & \VjepaTies{} & \VjepaUnstable{} & \VjepaHallucinations{} & 0 & \VjepaCoverage{} \\ \bottomrule \end{tabular}
\end{table}\begin{table}[!ht]
  \caption{Potential triage actions for outcome classes; benefits remain to be validated with developers.} \label{tab:actions} \centering \small \begin{tabular}{@{}p{0.25\columnwidth}p{0.68\columnwidth}@{}} \toprule Outcome & Suggested regression action \\ \midrule CP & Retain as a versioned baseline; flag any loss. \\ CDP & Track configuration dependence across serving or processor changes. \\ Inconclusive & Report the 2--2 split; never convert it into a pass. \\ CDF & Inspect event visibility, sampling, and scoring API behavior. \\ CF & Record a consistent failure; inspect the affected test category. \\ Incomplete & Inspect execution errors or abstentions. \\ \bottomrule \end{tabular}
\end{table}

Table~\ref{tab:routes} gives the primary SUT outcomes. Cosmos's modal accuracy is \CosmosAccuracy{} (95\% CI [66.4\%, 74.8\%]) and V-JEPA~2's is \VjepaAccuracy{} [44.1\%, 53.0\%]. Their failure profiles also differ. Cosmos's \CosmosHallucinations{}/\CosmosClassified{} consistent fails (\CosmosHallRate{}, [\CosmosHallLow{}, \CosmosHallHigh{}]) concentrate in test category P4 Solidity (\CosmosPfourCount{}/\CosmosHallucinations{}); it has \CosmosFragile{} CDP and \CosmosUnstable{} CDF tests. V-JEPA~2 has \VjepaHallucinations{} consistent fails (\VjepaHallRate{}, [\VjepaHallLow{}, \VjepaHallHigh{}]) spanning all categories. Across both SUTs, \ConfigurationSensitive{} of 875 decisive test outcomes (30.1\%) are configuration-dependent; \TotalTies{} inconclusive 2--2 tests also vary across settings. Verdict rate is \CosmosCoverage{} [76.1\%, 84.0\%] for Cosmos, with \CosmosTies{} inconclusive and \CosmosAbstainPairs{} incomplete tests, and \VjepaCoverage{} [90.5\%, 95.1\%] for V-JEPA~2. These profiles suggest different follow-up (Table~\ref{tab:actions}): inspect affected categories for CF, sampling and the scoring API for CDP/CDF, and missing evidence for non-decisive tests. This is a descriptive SUT contrast; model family, score semantics, and temporal interface are confounded, and triage benefit remains untested.\finding{\textbf{Answer to RQ2:} Cosmos has \CosmosFragile{} CDP and \CosmosUnstable{} CDF tests versus \CosmosHallucinations{} CF tests; V-JEPA~2 has \VjepaHallucinations{} CF tests. Outcome classes suggest different investigations that accuracy alone does not identify.}

\section{Implications and Threats}

\emph{Implications.} RQ1 implies that the test harness belongs inside the SUT boundary: preprocessing, temporal sampling, loss composition, model version, and serving code lie on the path from video to verdict and should appear in the build manifest \cite{sculley2015debt,pineau2021reproducibility}. The protocol needs only paired oracles, per-video scores, and configuration controls, so it also applies to black-box services. Each test costs four executions and records are append-only. This motivates regression test selection \cite{yoo2012regression}: a small smoke test suite on every serving change and the full benchmark on scheduled releases.

\emph{Construct validity.} Configurations need not carry equal evidence. IntPhys~2 lacks violation timestamps, so neither sampling nor sliding windows certifies event visibility; V-JEPA~2 context length changes both history and forecast horizon. Outcome transitions signal behavioral changes, not their cause. \emph{Internal validity.} Both revisions are post hoc case studies on one SUT, and the regression view was defined after observing them; they illustrate, but do not establish, detection power. \emph{External validity.} Results cover two SUTs and one synthetic benchmark; debugging benefit requires developer studies.

\section{Future Plans}

Our results suggest that regression testing of learned components needs oracles at the level of individual tests and configurations rather than aggregate scores. We plan four studies to turn this idea into a full paper and a shared research direction.

\emph{Harness faults in the wild.} Our padding fault was found by audit. We will mine open-source test-harness histories for faults in score conversion, sampling, padding, and aggregation, and measure how often they leave aggregate metrics nearly unchanged while altering per-test verdicts.

\emph{Oracle adequacy through mutation.} We will mutate score direction, pair order, frame sampling, loss aggregation, and model identifiers, and compare detection by accuracy deltas, negative-flip counts, configuration-level verdict diffs, and outcome diffs. As DNN adequacy measures can mislead \cite{li2019misleading}, we will test when outcome diffs justify four executions per test and which tests form an effective smoke test suite.

\emph{Attributing outcome changes.} Configuration-dependent behavior mixes model behavior with evidence availability. Annotated violation intervals, target-aligned histories, and fixed forecast horizons will help distinguish contributions from the model, harness, and test input.

\emph{Generality and practice.} The protocol needs only paired oracles, per-video scores, and configuration controls. We will cover more model versions and families, a second physics benchmark, natural videos, and non-video paired oracles. We will emit outcome diffs as pull-request reports in continuous integration and study with ML engineers whether they change release decisions compared with accuracy dashboards.

\emph{Open questions.} When does a test configuration set preserve semantics for learned components? How many configurations does an outcome oracle need, and how should transitions inform a release decision? We will keep the protocol, adapters, and reports open for additional SUTs, benchmarks, and revisions.

\section{Conclusion}

TetraState turns paired benchmark items into regression tests with an outcome oracle, explicit verdict rate, and a versioned SUT identifier. On two harness revisions, outcome diffs exposed changes that accuracy concealed, including \VjepaAblationHiddenRegressions{} regressions missed by a modal-verdict flip check. We envision outcome-transition reports accompanying revisions of learned components, as test-result diffs accompany code changes in continuous integration.

\section*{Acknowledgment}

\ifdefined\TetraStateThanks\TetraStateThanks\par\fi
\emph{Use of generative AI.} Claude (Anthropic) and OpenAI Codex assisted with code implementation and testing and with drafting and editing the text and figures. The research idea and study design are our own. We reviewed and validated all AI-assisted output, recomputed every reported number with the replication package, and take full responsibility for this work.

\section*{Data Availability}

An anonymous replication package (\ArtifactURL) provides the code, all test-execution records with raw scores, the official V-JEPA~2 loss artifacts, hash-bound run records, and a CPU-only script that recomputes every reported number and Fig.~\ref{fig:transitions}. IntPhys~2 videos and model versions are available from their original sources.

\bibliographystyle{IEEEtran}
\bibliography{references}

\end{document}